\documentclass[11pt,a4paper]{article}

\usepackage[top=15mm, bottom=15mm, left=15mm, right=15mm]{geometry}
\usepackage{color}
\usepackage[title]{appendix}
\usepackage{amsmath}
\usepackage{amssymb}
\usepackage{amsthm}
\usepackage{graphicx}
\usepackage{rotating}
\usepackage[shortlabels]{enumitem} 
\usepackage{wrapfig} 
\usepackage{multirow} 
\usepackage{cite}
\usepackage{placeins} 
\newcommand{\bdm}{\begin{displaymath}}
\newcommand{\edm}{\end{displaymath}}
\newcommand{\beq}{\begin{equation}}
\newcommand{\eeq}{\end{equation}}

\begin{document}
\bibliographystyle{abbrv}
{\Large \textbf{FROM MODELLING TO UNDERSTANDING:\\ THE SIGNALS IN NERVES}}\\
\\
\textbf{J\"uri Engelbrecht$^{[a]}$, Kert Tamm$^{[b]}$ and, Tanel Peets$^{[c]}$ }\\
\\
{\small
$[a]$  je@ioc.ee, Tallinn University of Technology, Department of Cybernetics and Estonian Academy of Sciences\\
$[b]$  kert.tamm@taltech.ee, Tallinn University of Technology, Department of Cybernetics\\
$[c]$  tanel.peets@taltech.ee, Tallinn University of Technology, Department of Cybernetics
}
\section*{Abstract}
This paper attempts to review our studies on the propagation of signals in nerves over the past decade.  The need for interdisciplinary studies is stressed that helps to understand the physical mechanisms of coupling the electrical, mechanical, and thermal effects in nerves. Based on the analysis of structural properties of axons and possible mechanisms of interaction between different physical phenomena, a set of assumptions and hypotheses is formulated. As a proof of concept, a rather general mathematical model is presented for describing a wave ensemble in unmyelinated axons. This model is composed of several governing equations (``building blocks") which are coupled by forces describing the interaction between the effects. The numerical simulation using the dimensionless variables demonstrated a rather good qualitative match with experiments. The further generalisation of this model in physical units for the processes in myelinated axons permits a closer match to measurements. Based on modelling and \emph{in silico} experiments, the guidelines for modelling such a complex electrophysiological process are formulated. These guidelines reflect the importance of following the physical principles in modelling together with interdisciplinary knowledge from continuum mechanics and mathematics.

\textbf{Keywords:} action potential, physical principles, mathematical modelling, interdisciplinarity

\section{Introduction}
The propagation of signals in nerves is an extremely important chapter of biology. This complex problem is intensively studied not only because of its possible pathological effects but also because of its role in cognitive processes and the processing of information. In what follows, attention is paid to the fundamental behaviour of signal propagation in healthy nerves under normal conditions focusing on physical phenomena. Numerous theoretical and experimental studies over the last two centuries have cast light on many details of this process. As demonstrated in experiments, the propagation of an action potential (AP) as a main carrier of information, is accompanied by mechanical and thermal effects. In addition to earlier experiments (see, for example, Terakawa \cite{Terakawa1985} and Tasaki \cite{Tasaki1988}), recent studies have also reported the mechanical response of mammalian neurons recorded by a label-free optical imaging method  \cite{Yang2018}  and high-speed interferometric imaging  \cite{Ling2020}. This means that there is an ensemble of waves and besides electrophysiology, attention must be paid to the complexity of the process involving physics, chemistry, and thermodynamics. In this context, mathematical modelling plays an important role because besides the experiments,  describing the physical mechanisms in nerves in mathematical terms helps to understand the causality and the coupling of effects. That is why studies of nerve propagation are at the interface of physics and mathematics.  

The need to turn attention to interdisciplinary studies in biology is stressed by many authors. Noble  \cite{Noble2002} stressed the need for integrative studies at all levels of biology while McCulloch and Huber  \cite{McCulloch2002} demonstrated how these ideas permitted to building up of an integrated cardiac model (see also Bassingthwaighte \cite{Bassingthwaighte1995}). The need for integrative studies in neurosciences was already known to Sherrington  \cite{Sherrington1907} who authored a monograph entitled ``The Integrative Action of the Nervous System". Winlow  \cite{Winlow2024}, however, claimed that ``...we often only model small parts of the multi-biophysics of cell membranes ... For the future, neuroscientists need to consider all the background physical and biophysical details and thei \cite{Hodgkin1964a} mentioned: ``In thinking about physical basis of action potential perhaps the most important thing to do at the present moment is to consider whether there are any unexplained observations which have been neglected in an attempt to make experiments fit into a tidy pattern''. Kaufmann  \cite{Kaufmann1989} has stated that ``electrical action potentials are inseparable from the force, displacement, temperature, entropy and other ... variables''. This is a challenge as emphasised by Andersen et al.  \cite{Andersen2009} -- there is a need ``.... to frame a theory that incorporates all observed phenomena in one coherent and predictive theory of nerve signal propagation.'' Despite the warning by Winlow  \cite{Winlow2024}, there are many attempts to construct mathematical models that involve not only the AP but also the accompanying effects \cite[etc.]{Jerusalem2014,ElHady2015,Chen2019,Schneider2021,Raamat2021}. These mathematical models take usually the AP as a driving force and use different assumptions for coupling one or another accompanying effect. In general, in addition to the AP, the accompanying effects include the longitudinal wave in the biomembrane (LW) and the corresponding transverse wave (TW), the pressure wave (PW) in the axoplasm, and temperature change ($\Theta$) accompanied by some biochemical changes. It can be said that summing up single components, an ensemble of waves is propagating in a nerve fibre  \cite{Engelbrecht2018c,EngelbrechtMEDHYP}.
 
In what follows, is an attempt to unify all the assumptions collecting the present knowledge for aiming to build up a basis for further studies in the sense of Andersen et al. \cite{Andersen2009}. We start here from the basic laws of physics that form the basis of the governing equations  \cite{Pennycuick1992} which are modified according to experiments in electrophysiology. These modifications are analysed paying attention to possible simplifications and coupling effects. Based on this analysis and the structural properties of nerve fibres, the main hypotheses and assumptions are formulated. As a result, a backbone of a mathematical model is constructed which is general enough to be specified when new evidence about the process is acquired. In addition, the mathematical background of modelling is analysed. Indeed, already Galileo Galilei said that the Book of Nature is written in the language of mathematics. The contemporary understanding of the importance of mathematics in biology is presented in a report of the National Research Council  of the United States of America \cite{NationalResearchCouncil2005}. Briefly, this Report says: ``... a mathematical model can highlight basic conceptions and identify key factors or components of a biological system''. It means that from modelling we enhance understanding. One of the complicated tasks is selecting the essential information from experiments and again, by casting the information into mathematical terms, the logic of mathematics helps to focus on key issues. 

Further on, we start in Section 2 with a brief review of the existing models based on our previous review  \cite{Peets2023}. Section 3 is devoted to basic principles that need to be taken into account in modelling. The next Section 4 describes the physical structure of an axon and the scales of its elements. Then the assumptions and hypotheses needed for modelling the signal propagation in nerves are presented in Section 5. This allows to construction of a mathematical model that demonstrates the correctness of basic ideas and since it is presented in the dimensionless form serves as a proof of concept. A more detailed analysis follows in Section 6 where attention is paid to the accuracy of assumptions. Finally, in Section 7, the conclusions are summarised. 

The article attempts to review our studies in a coherent framework and summarises the interdisciplinary ideas for modelling the signals in nerves.

\section{Brief review of recent mathematical models}

This review is based on our longer paper  \cite{Peets2023}. The attention is paid to recent mathematical models where the wave ensemble contains besides the AP one or more accompanying components. Most studies use the Hodgkin-Huxley paradigm \cite{Hodgkin1964a}, i.e., the AP is the primary component of an ensemble and triggers all the other effects.

A model of coupled electrical and mechanical effects based on the spring-dampers (dashpots) system is proposed by J\'erusalem et al. \cite{Jerusalem2014}. This model describes the process in a myelinated axon and the difference in the behaviour of the nodes of Ranvier and internodal regions is taken into account by the Hodgkin-Huxley model and cable theory. 

El Hady and Machta \cite{ElHady2015} have elaborated a mathematical model based on the assumption that the potential energy is stored in the biomembrane and the kinetic energy in the axoplasmic fluid. The model takes the AP without calculations as a Gaussian pulse and the attention is to determine the LW, TW, and $\Theta$. It is stated that the mechanical modes are driven by the changes of separation across the membrane. Although the PW is not described, it is assumed that its (called the bulk flow) influence is seen as the surface waves, i.e., waves in the biomembrane. The question is that according to the general understanding (see Malischewsky \cite{Malischewsky1987}), surface waves are depth-dependent and this property is not analysed. The heat is assumed to depend on summing up the amplitudes of LW and PW, however, a detailed analysis of such an assumption is not given. The profiles of the LW, TW, and $\Theta$ correspond qualitatively to the measured ones.

Chen et al.  \cite{Chen2019} proposed a coupled mechanoelectrophysiological model for axons that is based on using the flexoelectric effect. This means that changes (c.f. Section 2.1.5) of voltage field during an AP induce strain gradient fields on the axon resulting in a change of the membrane surface curvature (usually called the reverse flexoelectric effect). The AP is governed by the Hodgkin-Huxley model and cable equation, and both unmyelinated and myelinated cases are analysed. The biomembrane is taken as an elastic or viscoelastic cylinder with a thin wall and the conservation of momentum is used for deriving the corresponding model of mechanical effects in such a cylinder. This model includes the body force due to the flexoelectric effect. The change in the axon diameter is taken into account together with the changes in the membrane capacitance and resistance. The finite-element method is used for the numerical simulation and the calculated TW has a bipolar shape. The model permits the reciprocity of electrical and mechanical effects.

A mathematical model involving all the components of the wave ensemble (AP, LW, TW, PW, $\Theta$) is proposed by Engelbrecht et al.  \cite{Raamat2021}. This model is based on the Hodgkin-Huxley paradigm and the governing equations of all the components of the ensemble except the TW are derived from the basic principles and coupled by additional forces expressing the coupling mechanisms. The TW is related to the gradient of the LW. For modelling the effects from exo- and endothermic reactions to temperature changes, the concept of internal variables is used. In this model, the main attention is on modelling the accompanying effects, and therefore a simplified model for the AP -- the FitzHugh-Nagumo (FHN) model -- is used in most calculations, although it is possible to use also the standard Hodgkin-Huxley (HH) model or its modifications. The numerical simulation by the pseudospectral method demonstrated a good qualitative match with experimental studies.

In principle, there is one more theoretical possibility to generate signals in nerves besides the Hodgkin-Huxley paradigm. Rvachev \cite{Rvachev2010}  has assumed that the pressure wave PW in the axoplasmic fluid can trigger other phenomena in axons including the formation of an AP.

An analysis of these models gives ideas to elaborate general principles of modelling the signals in nerves with a better accounting of structural properties of nerve fibres which should enhance their predictive power.
Some models should be added to those described above because they help to specify certain components of the wave ensemble.
 
Schneider \cite{Schneider2021} stresses the importance of using physical principles to derive biological functions: He has studied experimentally and theoretically processes in monolayers that also permit a better understanding of the processes in bilayers. This makes it possible to unite electrical and mechanical pulses in lipids and state that the acoustic pulses in lipids have similarity to action potentials \cite{Mussel2021}.

A completely different idea for explaining the signals in nerves is proposed by Heimburg and Jackson  \cite{Heimburg2005} who consider a main signal as an ``electromechanical soliton''. This signal is a longitudinal wave of phase transition in the biomembrane and all the other phenomena in nerves are triggered by this mechanical wave. Although this model can describe the whole process from a different viewpoint compared with the Hodgkin-Huxley model, it is not clear how the electrical signal measured by numerous experiments, is formed. It is assumed that the membrane potential is linearly proportional to the density change  \cite{Heimburg2005} but this assumption does not explain the measured asymmetric shape of an AP (asymmetry means an overshoot) or the refractional overshoot (cf. the Hodgkin-Huxley \cite{Hodgkin1964a} or FitzHugh-Nagumo models \cite{Nagumo1962}). One should, however, note that the Heimburg-Jackson model is of great importance for explaining the dynamics in the general theory of cells where the phase transition may occur. However, it is not clear whether the nonlinear pulse generated in real conditions corresponds to the definition of solitons formulated in mathematical physics \cite{Ablowitz2011}.

Once different assumptions lead to different models, there is a need to unify the ideas of modelling for matching as well as possible observations. Several reviews must be mentioned  \cite{Jerusalem2019,Schneider2021,Drukarch2021,Carrillo2023} where various models were analysed. Still, the interdisciplinary ideas explaining how the knowledge from physics, chemistry, and mathematics is generalised into the framework of electrophysiology must be clearly formulated. The following presentation enlarges the scheme presented in the review by Peets et al. \cite{Peets2023} starting from general principles to the fully coupled model describing the wave ensemble and its possible modifications.

\section{Basic principles of modelling}

Before studying concrete problems, the basic considerations must be understood for a better understanding the natural phenomena. That is why we briefly discuss some ideas of interdisciplinarity needed for modelling complex processes.

The analysis of complex processes as seen from the philosophical viewpoint is presented by DeLanda \cite{DeLanda2002} who has introduced the notions of nonlinear dynamics for explaining ontological (existence, being, and becoming) and epistemological (knowing that) concepts.  As stressed by Holdsworth  \cite{Holdsworth2006}, DeLanda ``recovers for mathematical practice a capacity to clarify the meaning of events as they arise within a synthetic process of becoming interdisciplinary''. Some of DeLanda's ideas must be stressed.

\emph{General principles:}\\
- complex processes are characterised by multiplicity which is the activator or changes in the system;\\
- multiplicity is characterised by differences that are productive and cause interactions;\\
- the changes (gradients) are characterised by velocities;\\
- the causality for processes is related to multiplicities;\\
- emergence means a process where novel properties or capacities emerge from causal interactions.

\emph{For dynamical systems:}\\
- one should distinguish between intensive and extensive properties of systems: intensive properties like pressure, temperature, density, etc.~cannot be divided, extensive properties like length, area, volume, and amount of energy can be divided into parts, intensive properties have critical thresholds, differences (gradients) in intensity store potential energy;\\
- one should distinguish between intrinsic (belonging to the system) and extrinsic (originating from outside) conditions for a system;\\
- one should understand the inertiality of a system and the role of thresholds and triggers in dynamical processes;\\
- every physical process means also the transfer of information.

\emph{Finally:}\\
- for understanding complex processes, interdisciplinarity is needed.

These ideas are certainly universal in modelling of complex systems. Engelbrecht et al. \cite{PhilArtikkel} have analysed how DeLanda's formulations correspond to the general concepts for describing signal propagation in nerves. Next, few remarks about the physical background of the propagation of signals in nerves. Whatever the dynamical process in continua, it is governed by laws of physics. The modelling of waves starts from the conservation laws: electrical signals are governed by Maxwell equations, and mechanical waves -- by Newton's Second Law. In simple cases, the wave equation governs the process while external sources are modelled by additional forces in the governing equation. In the case of nerve signals, these additional forces are responsible for coupling the effects. The thermodynamic effects are modelled by the Fourier Law (heat flux is related to temperature gradient) and Joule's Law (heat is related to the electric current). The governing equations in electrophysiology are certainly modified to stress one or another specific property of the process but all possible simplifications must be carefully checked against the observations. One can say that physics shapes signals in nerves (Engelbrecht et al., 2022b) \cite{Peets2022}. The proper usage of laws of physics guarantees the consistency of models (see also Schneider, 2021) \cite{Schneider2021}.

However, not all the processes in electrophysiology are understood in detail, especially when several processes are coupled. Nevertheless, observations (experiments) help to understand many aspects of processes under investigation, which permits to description of empirical relationships between the phenomena. In this case, a model is based on phenomenology rather than on physical theory. Portides (2011) \cite{Portides2011} explains that such a model compensates for the lack of knowledge ``of how exactly and to what extent each part of a system contributes to the latter's investigated behaviour''. In other words, most of the system is too complex for a straightforward application of fundamental laws, which are typically composed of an idealised model system, and phenomenological models are used to describe the relationships between the variables of the model within the measured values.

The main idea of constructing phenomenological models is the following. The physical mechanism of changes of a certain variable is not known but its values can be measured. It means that from observations (experiments) one can estimate the initial and final values of a variable together with the time needed for such a change. Then a simple phenomenological model involving two parameters -- the amplitude change and the relaxation time -- is a first-order kinetic equation. The celebrated Hodgkin-Huxley model  \cite{Hodgkin1952} involves three phenomenological variables $m$, $n$, and $h$ that control the ion currents. In continuum mechanics, the notion of internal variables is used \cite{Maugin1990} for describing the dissipative processes but in essence, an internal variable is phenomenological. The usage of phenomenological variables in modelling of complex processes is in more detail analysed by Engelbrecht et al.  \cite{Engelbrecht2024}.

\section{Structure of axons}

The main structural element in nervous systems is the axon along which an electrical signal (action potential AP) propagates. The structure of axons and their morphology are studied in detail by Clay \cite{Clay2005} and Debanne et al. \cite{Debanne2011}. Nevertheless, here we present a brief overview with some explanations needed for modelling the physical processes in axons.

An axon can be modelled as a tube in a certain environment called an extracellular fluid. Inside the tube is the axoplasmic fluid (intracellular fluid) which contains cytoskeletal filaments. In terms of continuum mechanics, this fluid is viscous and has a microstructure. Both extra-- and intracellular fluids have a certain concentration of ions. The wall of the tube has a bilayered lipid structure called a biomembrane. It is composed of two layers of amphiphilic phospholipids with hydrophilic heads and hydrophobic tails. Again, in terms of continuum mechanics, it has a microstructure. Moreover, it is inhomogeneous because it contains the ion channels that play an important role in maintaining the steady shape of a propagating AP. The ion currents, i.e., the flow of ions through these channels regulate the shape of the signal by electrochemical gradients.

\begin{figure}[h]
\includegraphics[width=0.99\textwidth]{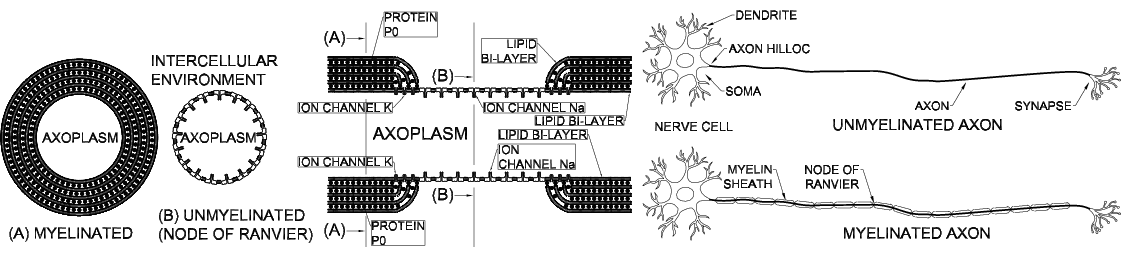}
\caption{Myelinated (A), unmyelinated (B) cross-sections of axon, node of Ranvier, and ion channels (C)}
\label{fig1}
\end{figure}

Actually, there are two types of axons: unmyelinated and myelinated. In the first case, the wall of the axon is just a single bilayer, in the second case this bilayer is covered by a myelin sheath which consists of multiple layers of a glial membrane composed of lipids and proteins. This sheath serves as an insulator for ion currents but is interrupted by so-called Ranvier nodes through which the ion change occurs. The classical experiment by Hodgkin and Huxley \cite{Hodgkin1952} was carried on  an unmyelinated axon. The existence of the myelin sheath (again a certain microstructure) causes certain changes in the AP velocity  \cite{Huxley1949} that must be accounted for in the modelling of the process.
The schemes of axons are presented in Fig.~\ref{fig1}.

In modelling, attention must be paid to the scales of structures and processes. The observations are summarised as follows \cite{Iwasa1980,Tasaki1989,Lodish2004,Ling2020}:\\
- Wavelengths of the propagating signals are essential for the modelling. If the signal duration is 2~ms and velocity is 2~m/s then the spatial length of the signal from start to finish is roughly 4~mm. \cite{Tamm2021}. If the duration is 2~ms and velocity 100~m/s then the spatial length of the signal from start to finish is roughly 20~cm. However, it must be noted that depending on the shape of the signal the spectral composition could include higher harmonics or frequency components with shorter wavelengths which could, in theory, be short enough to be sensitive toward smaller structures like ion channels or maybe even larger proteins. \\
- Axon diameter varies from a micrometer in certain nerves of the human brain to a millimetre in the giant fibre of the squid. Axon length varies from millimetres up to about a meter (giant fibre of the squid).\\
- The cycle of membrane depolarization, hyperpolarization, and return to the resting value that constitutes an action potential lasts 1--2~ms and can occur hundreds of times a second in a typical neutron.\\
- The node of Ranvier (a structure found on myelinated axons) is typically around 1~\textmu m in length (but may have lengths up to 50~\textmu m \cite{Tomassy2014})  and has a high density of ion channels. \\
- Myelinated segment is typically from $\approx$50 to $\approx$300~\textmu m in length.\\
- Lipid bi-layer is typically 3--4~nm in thickness.\\
- Mechanical transverse displacement during AP propagation is typically $\approx$1~nm in amplitude or less for mammalian neurons. However, the recent studies  \cite{Lefebvre2024} reported also the larger displacements in sub-nanometer range.\\
 - The AP can move down an unmyelinated  axon  at speeds up to 2~m/s (up to 25~m/s in squid giant axon) \cite{Purves2017}. In non-myelinated neurons, the conduction velocity of an action potential is roughly proportional to the square-root of the diameter of the axon.\\
- The presence of a myelin sheath around an axon increases the velocity of impulse conduction up to 120~m/s. Conduction velocity in myelinated axons is roughly proportional to the axon diameter.\\
- The myelin sheath is a stack of specialised plasma membrane sheets produced by a glial cell that wraps itself around the axon.\\

The scales listed above permit summarising the main features of nerve pulse propagation. Typical processes in time happen from microseconds (phase change of the lipid bi-layer in some of the described models) up to hundreds of milliseconds (temperature effects persisting after the nerve pulse has passed in some models) with most of the models predominantly focused on describing effects that are from a millisecond up to a few tens of milliseconds in time and, roughly, in phase with the main driving signal (AP or mechanical change in most of the described models). In the spatial resolution the noted models fall roughly into two broad categories, first, the models that are the most focused on the lipid membrane and the changes happening within (membrane thickness 3--4~nm, membrane displacement of roughly 1 nm but along the axon, the signal can be from millimetres up to tens of cm in scale) and second, the models that are taking a more ``continuum mechanics'' approach, focusing more on the macroscopic effects in spatial scales comparable to the length of the axon and sometimes including the influence of the smaller structures (like, for example, mechanosensitivity of some ion channels) in a roundabout way indirectly through the parametrisation of models.

\section{A possible set of assumptions and hypotheses}

Here we follow the following set of {\bf assumptions}  \cite{NovaPeatykk}:\\
\begin{itemize}
\item electrical signals are the carriers of information and trigger all the other processes  \cite{Hodgkin1952};
\item the axoplasm can be modelled as a fluid where a pressure wave is generated due to the electrical signal  \cite{Terakawa1985};
\item the biomembrane can deform (stretch, bend) under the mechanical impact  \cite{EngelbrechtTammPeets2014,Heimburg2005};
\item the ion channels in biomembranes can be opened and closed under the influence of electrical signals as well as mechanical input \cite{Mueller2014};
\end{itemize}

This means that we follow here the Hodgkin-Huxley paradigm. Although this paradigm has been criticised  \cite{Drukarch2018} as too restrictive (not including thermodynamic variables), there is no better model for describing the formation and propagation of the AP.
Later we explain that the proposed model is built in such a way that the generated AP may even be an experimentally measured one.

To explain the accompanying effects, the existence of interaction forces between the components of the signal coupled into a whole is needed. The physical mechanisms responsible for {\bf coupling} and generating an ensemble of waves are:
\begin{itemize}
\item electric-biomembrane interaction resulting in mechanical waves (longitudinal -- LW and transversal -- TW) in the biomembrane;
\item electric-fluid (axoplasm) interaction resulting in a mechanical wave in the axoplasm (PW);
\item electric-fluid (axoplasm), electric-biomembrane, and mechanical-biomembrane interaction resulting in a thermal response ($\Theta$) in the fibre.
\end{itemize}

A detailed analysis of observations and mechanisms of interaction between all components
of a signal is presented by Engelbrecht et al.  \cite{Engelbrecht2020m,Raamat2021}. Based on this analysis, the following {\bf hypotheses} are made  \cite{EngelbrechtMEDHYP,Raamat2021}:
\begin{itemize}
\item  field variables (a components of a signal) are influenced by changes in other field variables;
\item the interactions are modelled by the coupling forces between the components of a signal;
\item all mechanical waves in the axoplasm and the surrounding biomembrane together with the heat production are generated due to changes in electrical signals (AP or ion currents) that dictate the functional shape of coupling forces; in mathematical terms, changes are described by derivatives of field variables;
\item the formalism of internal variables can be used for describing the exo-- and endothermic processes of heat production;
\item not only the influence of an AP on other effects but also possible feedback is considered.
\end{itemize}

\begin{figure}[h]
\begin{center}
\includegraphics[width=0.5\textwidth]{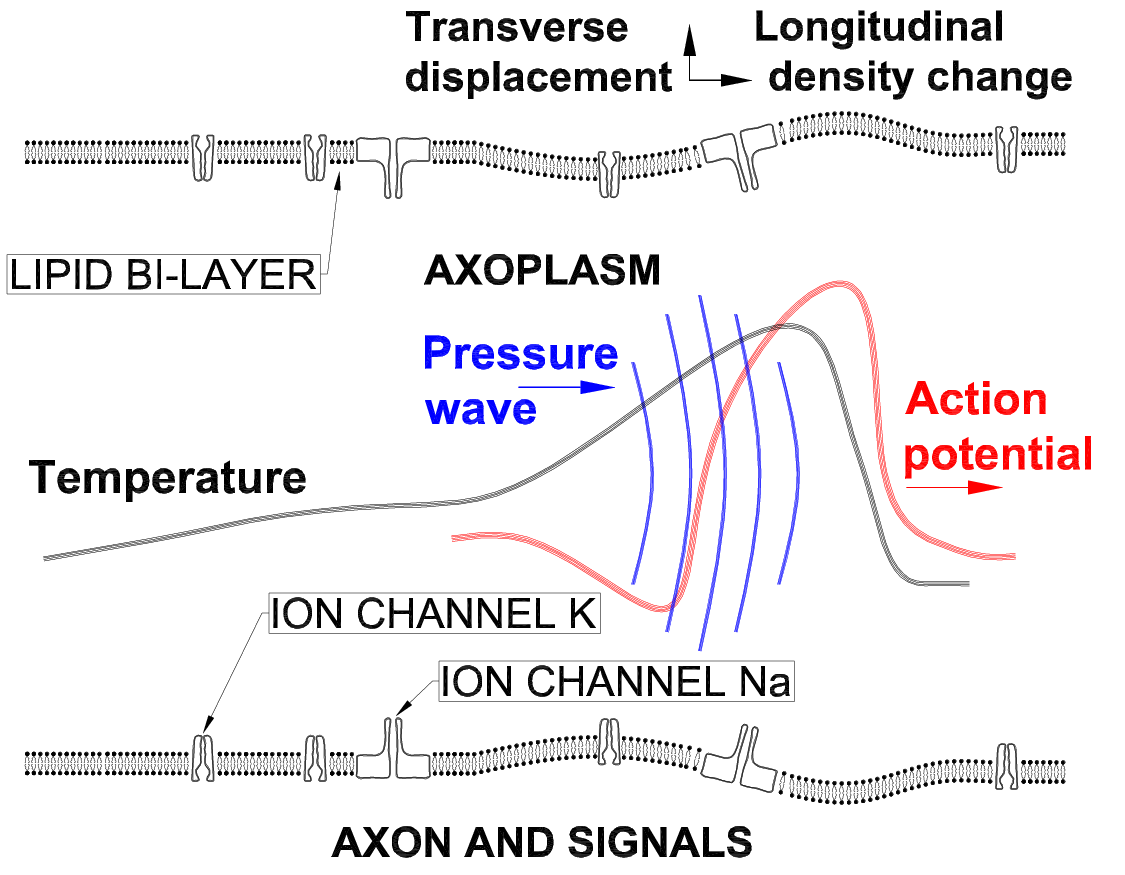}
\end{center}
\caption{Sketch of a wave ensemble in an axon.}
\label{fig2}
\end{figure}

The first hypothesis is related to the Du Bois-Reymond Law: ``The variation of current density, and not the absolute value of the current density at any given time acts as a stimulus to muscle or motor nerve'' (cited after Hall  \cite{Hall1999}) and the variation means mathematically the derivatives of field variables. Consequently, the possible mechanisms of coupling are described by space or time derivatives of variables governing the AP (amplitude $Z$), ion current $J$ (or $J_K$ and $J_Na$ ), the LW (amplitude $U$), PW (amplitude $P$), and temperature $\Theta$. These derivatives are related to the mechanisms of interaction listed above (see detailed description by Engelbrecht et al. \cite{Raamat2021}). The transverse displacement of the biomembrane $W$ is taken proportional to the gradient of $U$ as in the well-elaborated theory of rods \cite{Porubov2003}. This relationship is usual in continuum mechanics but it needs to be also understood in electrophysiology. Considering all the effects, it means that a signal in a nerve is an ensemble of waves depicted in Fig.~\ref{fig2}.

Based on assumptions and hypotheses, it is possible to build up a mathematical model that describes all the elements of the ensemble. In the first stage, for the sake of generality, we use dimensionless variables and start with modelling the processes in an unmyelinated axon. It permits us to pay attention to coupling effects. The model is a system of coupled differential equations and every ``building block" in this system of equations can be replaced with a better one if this is available. Even an experimentally measured AP could be used, only, in this case, the feedback from other effects on the AP is lost.

The model consists of the following parts:\\
\textbf{(i)} The action potential (AP) is modelled either by the FitzHugh-Nagumo (FHN) model  \cite{Nagumo1962} or by the Hodgkin-Huxley (HH) model \cite{Hodgkin1952}. The FHN model includes only one abstracted  ion current and is capable of modelling the main properties of the AP. If the effects of individual  ion currents need to be considered, then the HH or its modifications can be used.\\
\textbf{(ii)} The pressure wave (PW) in axoplasm is modelled by a wave equation with viscous and coupling terms. The coupling terms model the electric-fluid interaction. Depending on the parameters of the coupling force, the PW might have also a bipolar shape like demonstrated by Terakawa  \cite{Terakawa1985}.\\
\textbf{(iii)} The longitudinal wave (LW) in biomembrane is modelled by the improved Heimburg-Jackson (iHJ) model  \cite{EngelbrechtTammPeets2014,Heimburg2005} with coupling terms. The iHJ model accounts for the elasticity and inertia of the embedded lipid structure of the biomembrane The coupling terms model the electric biomembrane interaction and possible interaction with the PW.\\
\textbf{(iv)} The transverse displacement (TW) is calculated from the LW taking it proportional to the gradient of $U$ like in the theory of rods (Porubov, 2003) \cite{Porubov2003}. The TW has typically a bipolar shape.\\
\textbf{(v)} The thermal response $\Theta$ is governed by the classical heat equation with coupling terms. Coupling terms arise from the Joule heating, dissipation from the mechanical waves, and possible exo- and endothermic effects. 

The formalism of internal variables  \cite{Engelbrecht2020a,Raamat2021,Maugin1990} is used for modelling the exo- and endothermic effects. 
Consequently, the ensemble that is composed of AP, PW, LW, TW, and $\Theta$, can be divided into primary and secondary components \cite{Engelbrecht2020bmmb}. The primary components are characterised by corresponding velocities and their mathematical models are described by models supporting wave-like solutions. These components are the AP, PW, and LW. The secondary components are either derived from the primary components (like the TW) or their models are derived from basic laws that do not involve the velocities like the diffusion type equations governing temperature $\Theta$. In general terms, the ensemble of waves is formed due to changes in the intrinsic variables of the system. 

Note that whatever the models for an AP are, its main features need to be taken into account:\\
-- the existence of a threshold for an input;\\
-- the all-or-non phenomenon for a pulse;\\
-- the existence and the propagation of an asymmetrically localised pulse with an overshoot, i.e., the existence of a refraction length;\\
-- the possible annihilation of counter-propagating pulses.\\   
The model is described in Appendix 1. The computational results (see Fig.~\ref{fig3}) demonstrate a good qualitative match with experimentally measured profiles.

\begin{figure}[h]
\includegraphics[width=0.44\textwidth]{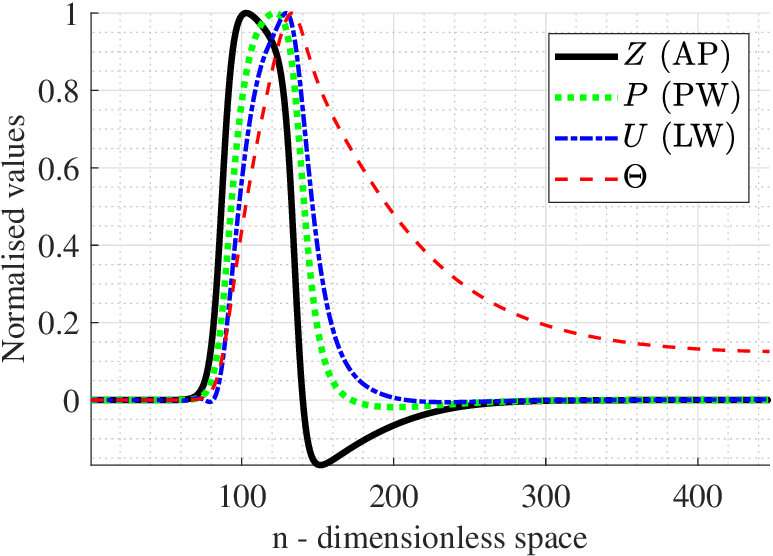}
\includegraphics[width=0.48\textwidth]{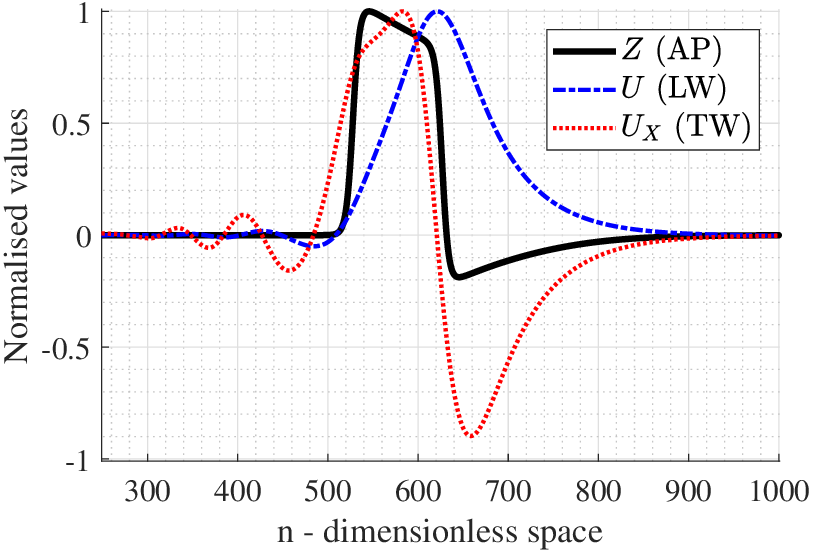}
\caption{The wave ensemble (from Engelbrecht et al 2021b) \cite{Peets2022} in an unmyelinated axon:
          (a) AP, PW, LW, $\Theta$ and
          (b) LW and TW}
\label{fig3}
\end{figure}

\section{Discussion}

Based on general principles of modelling physical processes (Section~3), the modelling of signals in nerves is described in Sections~4 and 5 (see also the Appendix). In deriving the mathematical model, one should keep some useful ideas in mind. First: ``Everything should be made as simple as possible but not simpler'', as stressed by Albert Einstein \cite{Robinson2018}. Second, the formation of a wave ensemble in nerves is a nonlinear complex process and we shall follow the advice of Gerald Whitham  \cite{Whitham1974} who recommended paying attention to small terms ($\varepsilon$-type) in governing equations that may influence the process in the long run. 

In terms of the theory of mathematical modelling, the proposed model in Section~5 is of a hybrid character  \cite{Morrison1999} involving besides physical laws phenomenological variables. The model is composed of parts describing different physical phenomena coupled with employing contact forces. In some sense, this model is a proof of the concept. At the first stage, the AP is described by the simplest possible FHN model, possessing just one abstracted ion current. The AP calculated by the FHN equation has a correct shape, needed for generating other accompanying components in the wave ensemble. The idea of this model involves the possibility of replacing every single element if a better understanding of the process is available. Later in Appendix, we demonstrate using the modified HH model  \cite{Lieberstein1967} for calculating the AP in the case of a myelinated axon. Although the standard HH model is widely used, there are also proposals to account for different ionic currents. Morris and Lecar \cite{Morris1981} proposed a model with Ca$^{2+}$  ions while Deng \cite{Deng2017} considered a different molecular mechanism of K and Na ions resulting in a simpler governing equation (instead of $n^4$ and $m^3h$, only $n$, $m$, and $h$ were responsible for changes in ion concentration). The cases when these proposals are justified should be verified by experiments.
  
The propagation of signals in nerves is an extremely important chapter in neurophysiology and the proper modelling of the physical background of processes could help to understand better neuronal activities. It is quite clear that the models undergo tight analysis. For example, even the HH model is criticised for not being able to describe accompanying effects  \cite{Drukarch2018} although it is clearly stated by the authors that it describes only electrical effects. Concerning the coupled model involving electrical, mechanical, and thermal effects (Section~5), the problem of consistency is raised  \cite{Holland2019}. However, for this model, it is shown that even at the starting phase, the governing equations are based on physical laws  \cite{Peets2022} and then one should ask about the consistency of conservation laws in general. Once the governing equations are modified then the assumptions made for simplifications should also be critically analysed. It must be stressed again that the model described above served first for establishing proper and physically grounded principles for constructing its elements and all the ``blocks'' in it could be replaced by better descriptions. Unfortunately, despite many analytical (and critical) overviews, no better models have been proposed.

Returning to the analysis of various assumptions made so far for deriving the models (including the model described in Section~5), some remarks are in order. The influence of the change in the diameter of an axon during the propagation of waves might indeed cause changes in other physical properties, However, the measured changes of the transverse displacement of the biomembrane (TW) are of the order of 1-2~nm  \cite{Tasaki1988,Tasaki1989}. The diameter of nerve fibres varies from 0.5~\textmu m to 25~\textmu m \cite{Lodish2004} up to 1~mm in the case of a  squid \cite{Hodgkin1952}. Consequently, according to Terakawa experiments which also used the giant squid, the area of the axon was changed ca 0.4\%. The membrane capacitance may be taken proportional to the surface area and then, given the diameter of mammalian axons up to 25~\textmu m, it means a change of the surface area by 0.1-0.2 \%. For a normal axon, these small changes might be neglected as Hodgkin and Huxley \cite{Hodgkin1952} neglected the inductance in their model. However, if more structural details are accounted for, like for example, the influence of the cytoskeleton then such small changes may play some role \cite{Whitham1974}. In the model of Chen et al. \cite{Chen2019} the change in the axon diameter is taken into account together with the changes in the membrane capacitance. Later in Appendix, the influence of these changes is accounted for together with the small inductance for modelling the AP in the case of a myelinated axon using the physical units.  

Heimburg and Jackson \cite{Heimburg2005,HJ2007} proposed a model of a signal in nerves based on the density excitation in the biomembrane, i.e., in the wall of a nerve fibre. This signal is a solitary wave and indeed, the corresponding governing equation (HJ equation for short) which is of the Boussinesq-type, possesses a solitary-type solution. Later Engelbrecht et al.  \cite{EngelbrechtTammPeets2014} improved this equation following the analysis of deformation waves in media with the microstructure. If the HJ equation includes only the influence of elasticity of lipid molecules constituting the biomembrane then the iHJ equation accounts also for the inertial properties. This means that the velocity of a signal is bounded for all frequencies. The nonlinearities are assumed to be of the displacement type contrary to the usual deformation-type nonlinearities in solid mechanics. The question is whether the main signal that carries information along the nerve is a soliton or not. In mathematical physics  \cite{Ablowitz2011}, a soliton is a solitary wave that maintains its shape, propagates with a constant velocity, and restores its shape and velocity after collision with another soliton except for the phase shift. Moreover, the velocity of the soliton depends upon its amplitude. The solution of the HJ or iHJ equations is indeed a soliton -- a symmetric $\text{sech}^2$ pulse. The question is about the physical properties of the biomembrane. According to data presented by Heimburg and Jackson 
 \cite{HJ2007}, the velocity of a density wave as a sound wave may be as high as 176.6~m/s and the other estimations indicate the value about 100~m/s. From the soliton theory, it is known that depending on the energy of the initial condition, the soliton train (the sequence of solitons) forms. The calculations by Tamm and Peets  \cite{Tamm2015} demonstrated that soliton trains are formed after 100-300~ms. It means then that the solitons form after propagating several meters which raises the question of whether it does correspond to physical case. Another and maybe the crucial question is, how the deformation wave in the biomembrane generates the AP? Up to now, there is no clear explanation for this mechanism. However, the iHJ equation may describe deformation processes in biomembranes and is relevant in cell mechanics.

Hodgkin and Huxley  \cite{Hodgkin1952} proposed to use of phenomenological variables for describing the ion currents. In the model proposed by Engelbrecht et al. \cite{Peets2021}, similar variables are used for describing the temperature changes due to endo-- and exothermic relations. The governing equations in both cases demonstrate the existence of relaxation processes. The calculations using a variant of the HH model \cite{Tamm2024} show that the relaxation time of phenomenological variables is in phase with the AP, which is about 20ms. Although in the dimensionless model, the temperature changes match qualitatively pretty well the measured profiles, the lack of physical parameters does not permit to calculate the relaxation time in physical units.

The physical basis for the electrical signals stem from Maxwell equations but in the celebrated HH model  \cite{Hodgkin1952} the inductance is neglected. Theoretically, it is clear that inductance is related to the velocity of the electrical signal. The experiments demonstrate that the HH model based on the ionic mechanism describes the process pretty well. However, for myelinated axons where the velocity under the myelin sheath is changing without the influence of the ion currents, not all the effects are understood. To grasp all possible effects, one might restore the neglected terms in governing equations to have a basis for comparing the models. This idea is supported by Wang et al. \cite{Wang2021} who have argued that inductance is ``a missing piece of neuroscience''. For the AP in unmyelinated axons, such a full model is proposed by Lieberstein  \cite{Lieberstein1967}. 

The analysis of processes in myelinated axons has revealed that under the myelin sheath, the velocity of the AP is increasing. This effect was already reported by Lillie \cite{Lillie1925} and is now referred to as saltatory conduction \cite{Huxley1949}. In the context of the full model, the question is not only the AP propagation but also in mechanical and thermal effects. The myelin sheath is a multilayered structure composed of glial cells that wrap around the axon  \cite{Lodish2004}. The sheath is interrupted by Ranvier nodes where the single bilayer like in unmyelinated axons acts as a region for ion currents. The segments of the myelin sheath are usually 50-300 \textmu m in length while the length of Ranvier nodes is about 1 \textmu~m or longer \cite{Lodish2004}. In terms of continuum mechanics, the structure of the myelin sheath can be compared with microstructured solids. Following the ideas of continuum mechanics, Tamm et al. \cite{Tamm2021} have proposed to model the longitudinal deformation in a myelin sheath by describing the complicated structure with an additional internal variable. The mathematical model is a system of two equations: the iHJ equation and the additional wave equation that governs the internal variable. In principle, this model can describe the changes in velocities and wave profiles in a myelin sheath but needs experimental verification. 

\begin{figure}[h]
\includegraphics[width=0.90\textwidth]{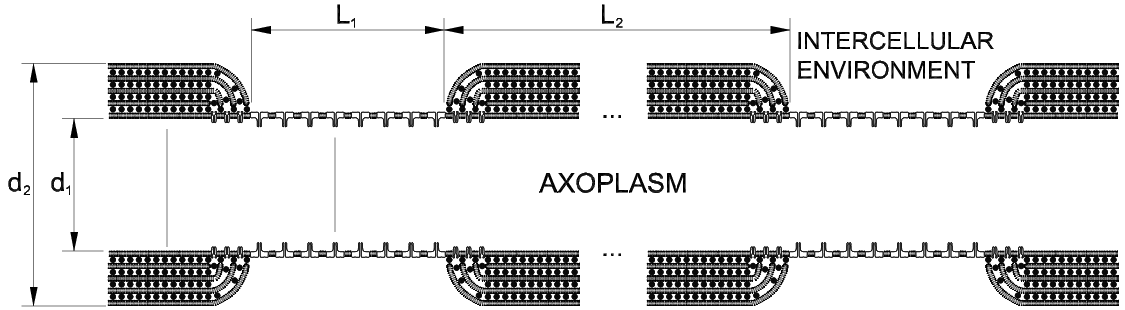}
\caption{The scheme of the myelinated axon.}
\label{fig4}
\end{figure}

According to the HH paradigm, the AP is a trigger for all the other effects in axons. The basic processes in unmyelinated axons (Section~5) can be modelled with  good accuracy, but the modelling of the AP in myelinated axons needs special attention. The scheme of a myelinated axon is shown in Fig.~\ref{fig4}.

It is proposed \cite{Goldman1968} that under the myelin sheath, the passive cable equation (the diffusion-type equation) governs the process and in Ranvier nodes, the usual HH equation works. Tamm et al.  \cite{Tamm2024} have specified the model by taking into account the structural properties of myelinated axons. It is proposed that the AP is governed by the Lieberstein-type \cite{Lieberstein1967} model where the inductivity (as small or large as it is) \cite{MelvinLieberstein1970} is kept as it follows from the basic Maxwell equations. The governing equation includes also the dependence of the capacity on the radius of the axon. After checking the validity of the model for an unmyelinated axon, the phenomenological dependence on the geometrical structure of a myelin sheath is introduced. This means introducing the g-ratio (the ratio of the outer vs inner diameter of an axon (see Fig.~\ref{fig4})) and \textmu -ratio (the ratio of the length of the myelin segment over the length of the Ranvier node (see Fig.~\ref{fig4})). In other words, the myelination geometry of across the axon (that is g-ratio) and along the axon (that is \textmu -ratio ) is taken into account. This assumption is partly based on physical considerations treating the myelinated segment as a classical Ohmic resistor  \cite{Bressloff2014}, but partly on phenomenology (\textmu -ratio). Using the data for the HH model  \cite{Lieberstein1967}, the calculations in physical units  \cite{Tamm2024} demonstrate the increase of the AP velocity up to 67.6~m/s depending also on the radius of the axon (from 1~\textmu m to 32~\textmu m) while the range of \textmu - ratio is from 0 (unmyelinated axon) to 325.    

The governing equation describing the propagation of an AP in the myelinated axon is presented in Appendix~2.

\section{Final remarks}

The modelling of processes in nerves is based on the structural properties of nerves and observations. The recent experimental results have demonstrated that electrical signals as main carriers of information, are accompanied also by mechanical and thermal effects. Some of the recent models were briefly analysed in Section~2. Our earlier studies were summarised in a monograph by Engelbrecht, Tamm, and Peets \cite{Raamat2021} but lately, attention has been paid to the general background of mathematical models from the view of interdisciplinarity. This review generalises our approach and several guidelines are stressed for building a solid basis for mathematical models.

\textbf{Guideline 1:} Physics rules electrical, mechanical and thermal biological processes.\\
The laws of physics are universal and serve as a basis for all processes in matter. The important notions in the context of nerve signals are electric current, mechanical deformations, temperature, and heat. Electrical currents are governed by Maxwell equations, although in many models the governing equation is modified or simplified (cf Hodgkin and Huxley \cite{Hodgkin1952}). Dynamical mechanical processes are governed by momentum conservation (usually referred to as Newton's Second Law). This leads to wave equations that often are modified by adding dissipative or dispersive effects. The thermodynamical effects are governed by the Fourier law (heat flux is related to the temperature gradient) and the Joule's law (heat is related to the electric current). The existence of additional forces means the influence of other fields and this serves as a basis for modelling the coupling between the single waves in an ensemble. The axioms of continuum physics (mechanics) should be followed (especially the axiom of equipresence). A more detailed analysis of applying physical ideas in biology is presented in overviews by Pennycuick (1992) \cite{Pennycuick1992}, Schneider (2021) \cite{Schneider2021}, and Engelbrecht et al. (2022b) \cite{Peets2022}.

Historically it is interesting to note that already Leonardo da Vinci (1452--1519) knew that one should ``observe the phenomenon and list quantities having numerical magnitude that seems to influence it'' (cited after Truesdell \cite{Truesdell1965}). In other words, one should first understand the basics of phenomena and after that proceed to details.

\textbf{Guideline 2:} Changes in one variable (field) will cause changes in other variables (fields).\\
The signals in nerves constitute a wave ensemble where single processes are coupled to each other. The coupling means that every change in one of the variables (fields) will cause changes in other variables (fields). This is actually the generalisation of the Du Bois-Reymond Law (see Section 5~and  \cite{Hall1999}). The physical hypothesis has an important consequence for mathematical modelling: the changes in variables are derivatives. It means that the coupling forces should involve the derivatives of variables either with respect to time or space. Combined with the physical considerations about the mechanisms of interaction (see Section~5) this forms the basis for describing the interaction processes.  

\textbf{Guideline 3:} Modelling of electrophysiological processes needs interdisciplinary studies.\\
Besides numerous experiments which have revealed the ``anatomy'' of nerve signals (see, for example, \cite{Clay2005,Debanne2011}), knowledge from other scientific disciplines may considerably enhance the understanding of processes in nerves. Indeed, there is the need ``to unravel the complexity of biological processes'', stressed by Noble \cite{Noble2002a} who suggested modelling in an integrative way. The philosophical considerations  \cite{PhilArtikkel} demand that one should distinguish between intrinsic (belonging to the system) and extrinsic (originating from outside) conditions and also between intensive and extensive properties. These considerations help better understand the dynamical processes. The knowledge of continuum mechanics helps to model mechanical effects in nerves either in the biomembrane or in the axoplasm. Heimburg and Jackson  \cite{Heimburg2005} derived their model for LW in biomembrane following the rules for deriving the wave equations and Engelbrecht et al. \cite{EngelbrechtTammPeets2014} improved this equation by distinguishing the elastic and inertial effects of the embedded phospholipid molecules like a microstructure. The idea of modelling the general effects of microstructure as an additional field permitted also modelling the mechanical waves in myelin sheaths \cite{Tamm2021}. The modelling of temperature changes as well as the pressure wave in the axoplasm follows directly the physical considerations. In terms of continuum mechanics, the axon is a long tube. It is well known that the longitudinal and transverse deformations of the tube wall are related. More precisely, the transverse displacement is proportional to the longitudinal deformation. It means that if the longitudinal displacement is a unipolar pulse then the transverse displacement is bipolar exactly as experiments demonstrate  \cite{Tasaki1988}. Interdisciplinarity means also that the terms should be used correctly. For example, in mathematical physics, the term ``soliton'' is precisely defined and the usage of this term is justified only after checking the properties of the solution.

\textbf{Guideline 4:} Phenomenology helps considerably to understand measurements.\\
Edmund Husserl (see \cite{Heelan1991}) already stressed that analysis of phenomena helps to understand the world. As described by Rovelli  \cite{Rovelli2022}, Heisenberg's idea to model the quantum world by matrices was just a phenomenological description that helped the further understanding of physical processes. In electrophysiology, the celebrated HH model involves three phenomenological variables $n$,$m$, and $h$ which are used for describing the ion currents  \cite{Hodgkin1952}. In continuum mechanics, such variables are called internal  \cite{Maugin1990}. Based on ideas of continuum mechanics, the internal variable is proposed by Engelbrecht et al.\cite{Engelbrecht2020a,Tamm2019} for describing the temperature changes in axons due to endo-- and exothermic reactions. The influence of the internal microstructure on processes in the main body can also be described by an additional field that in principle is a phenomenological approach. This approach is used by Tamm et al.\cite{Tamm2021} for modelling mechanical waves in the myelin sheath. For describing the propagation of the AP under the myelin sheath, also a phenomenological approach is used by introducing the $\mu$-factor which is a ratio of the lengths of a myelin segment and the Ranvier node.

\textbf{Guideline 5:} The experiments in electrophysiology serve as a basis for theoretical models.\\
The importance of observations in the context of explaining the physical basis of processes in nerves was mentioned already by Hodgkin  \cite{Hodgkin1964}. Mathematical modelling and \emph{in silico} experiments are powerful tools to reach a better understanding of signals in nerves. The \emph{in silico} experiments permit to cover of a large area of physical parameters in order to find suitable sets verified by experiments \emph{in vivo} or \emph{in vitro}. However, one should be aware of certain differences between physics and biology  \cite{Bialek2018}. He argued that in biology, there is a tradition to answer most questions by experiments while in physics, theory and experiments are more equal partners. In this context, biology is moving closer to physics.

The model proposed in Section~5 (see also Appendix) is built by coupled equations and in some sense, is composed of ``building blocks''. Every single block  (model equation) can be changed by a better one if such a governing equation can be improved on the basis of experiments. The computed profiles of the wave ensemble using this model as a proof of concept, have demonstrated good qualitative match with experiments. However, many problems need further studies to improve modelling. The improved HH model can describe the propagation of the AP for unmyelinated and myelinated axons \cite{Tamm2024} but improving the modelling of temperature changes needs more experiments for determining the parameters of governing equations in physical units. The further modifications of models (``building blocks'') should turn to details like accounting for various ionic currents, distribution of ionic channels. The influence of the cytoskeleton, membrane proteins, and the cellular structure in general, etc. An extremely interesting question is how the generation of an AP in the axon hillock generates the other elements of the wave ensemble. From the practical viewpoint, it is important to analyse the cases when one or another physical mechanism is ``out of order'' resulting in dysfunction of a nerve.                                                                                                                                                                  

Following the guidelines, listed above, the modelling is kept on a solid physical basis including also phenomenology supported by observations. In addition, philosophical considerations and clear terminology are useful in formulating the general framework of models. This helps the usage of interdisciplinary ideas. 

Finally, the idea of this paper was to sum up basic knowledge about the physics of signals in nerves and formulate the primary principles (guidelines) needed for modelling (and understanding) the propagation of signals. The next step is to improve the description of the model (governing equations) for a better accounting of the structural details of nerves and possible dysfunction.

\section*{Acknowledgements} 
  This research was supported by the Estonian Research Council (PRG 1227). Jüri Engelbrecht acknowledges the support from the Estonian Academy of Sciences.      

\begin{appendices}

\section{Dimensionless model}

Here, we briefly present the mathematical details of the model in dimensionless form \cite{Raamat2021}. Here subscript $X$ denotes partial derivative by $X$ (dimensionless space) and $T$ partial derivative by $T$ (dimensionless time). 

\begin{enumerate}[(a)]
\item
$
Z_{T} =  D Z_{XX} - J + Z \left( Z - \left[ a_1 + b_1 \right] - Z^2 + \left[ a_1 + b_1 \right] Z \right); \\
J_{T} =  \varepsilon \left( \left[ a_2 + b_2 \right] Z - J \right),
$
\item
$
P_{TT} = c_{f}^{2} P_{XX}  - \mu_1 P_T + F_1(Z,J,U),
$
\item
$
U_{TT} =  c_{0}^{2} U_{XX} + N U U_{XX} + M U^2 U_{XX} +
   N U_{X}^{2} + 2 M U U_{X}^{2} -
   H_1 U_{XXXX} + H_2 U_{XXTT} - \mu_2 U_T + F_2(Z,J,P),
$
\item
$
\Theta_{T} = \alpha \Theta_{XX} + F_3(Z,J,P,U),
$
\end{enumerate}
where $F_1 =  \eta_1 Z_X + \eta_2 J_T + \eta_3 Z_T$; $F_2 =  \gamma_1 P_T + \gamma_2 J_T - \gamma_3 Z_T$ and
$F_3 = \tau_1 Z^2 + \tau_2 \left( P_T + \varphi_2(P) \right) + \tau_3 \left( U_T + \varphi_3(U) \right) - \tau_4 \Omega$.

Here the AP (a) is governed by the FitzHugh-Nagumo (FHN) model \cite{FitzHugh1961} where $b_i = -\beta_i U$ and $Z$ is the action potential, $J$ is the ion current, $\varepsilon$ is the time-scales difference parameter, $a_i$ is the ``electrical'' activation coefficient, $b_i$ is the ``mechanical'' activation coefficient and $U$ is the longitudinal density change from lipid bi-layer density model (c) and finally, $\beta_i$ is an coupling coefficient. For the pressure wave in axoplasm (b) we use the classical wave equation with dissipation and force terms where $P$ is pressure, $\mu_1$ is viscosity coefficient and $c_{f}^{2}$ is the sound velocity in axoplasm. For the density change of the biomembrane (c) we use improved Heimburg-Jackson model \cite{Engelbrecht2018,HJ2007} where $U = \Delta \rho$ is the longitudinal density change, $c_{0}$ is the sound velocity in the unperturbed state, $N, M$ are nonlinear coefficients, $H_i$ are dispersion coefficients and $\mu_2$ is the viscous dampening coefficient, here $H_1$ accounts for the elastic properties of the bi-layer and $H_2$ the inertial properties. Finally, for the temperature (d) we use the classical Fourier law with the coupled source/sink term where $\Theta$ is the temperature and $\alpha$ is the thermal conductivity coefficient. The transverse displacement of the biomembrane is found through the membrane density change as $W \propto U_X$ \cite{Porubov2003,EngelbrechtTammPeets2014}.

The coupling terms: 
\begin{enumerate}
\item For $F_1$ (in the pressure (b)) the $\eta_i$ are coefficients and term $Z_X$ account for the presence of charged particles in the presence of potential gradient (along the axon), term
$J_T$ accounts for the ionic flows into and out of axon (across the membrane) and term
$Z_T$ accounts for the possible pressure change as a result of membrane tension changes from electrical field.
\item For $F_2$ (in the improved Heimburg-Jackson model (c)) the $\gamma_i$ are coefficients and term $P_T$ accounts for possible membrane deformation because of pressure changes (pressure to TW to LW), term
$J_T$ accounts for the possible membrane deformation as a result of ionic flows through ion channels and 
$Z_T$ accounts for the possible electrically induced membrane tension change. Note the sign, assuming that if tension increases then density decreases. 
\item For $F_3$ (in the Fourier law (d)) the $\tau_i$ are coefficients and term $Z^2$ accounts for the Joule heating, terms $P_T$, $U_T$ accounts for reversible temperature change (if density increases the temperature increases and vice versa), terms $\varphi_2(P)$, $\varphi_3(U)$ accounts for irreversible temperature change (energy lost to dissipation ends up as heat) and finally $\Omega$ characterises an abstracted endothermic chemical reaction working on a slower timescale than driving signals \cite{Tamm2019}.

\end{enumerate}

\section{AP in a myelinated axon}

Starting with the elementary form of Maxwell equations and drawing inspiration from the classical HH paper \cite{Hodgkin1952} it is possible to derive the model equations for unmyelinated axon similar to the Lieberstein model \cite{Lieberstein1967}. While Lieberstein opted to go a step further by moving into a moving frame of reference it is easier to solve by staying with the form which is closer to the Maxwell equations for a transmission line eqs.~\eqref{LIB1}, \eqref{LIB2} and just use the work of Lieberstein as a source of inspiration for handling the question of inductance $L$ in the context of signal propagation along the nerve axon. The governing equations demonstrate that the behaviour of the solutions is in the physiologically plausible range and the key characteristics of the nervous signalling are fulfilled \cite{Tamm2024}. These are: (i) the annihilation of AP signals during a head-on collision, (ii) the existence of activation threshold, and (iii) the refraction period after signal passing. In the parameter range considered, we observe the AP signal propagation velocity $c_{AP}$ from 0.5 $[\mathrm{m/s}]$ up to about 4.4 $[\mathrm{m/s}]$ for the unmyelinated axon \cite{Tamm2024}. The key difference between the classical HH model and the Lieberstein-inspired model used here is that the mechanism for signal propagation along the axon emerges more `naturally' as a consequence of opting to keep the inductivity $L$. While, indeed, there exist variations of the classical HH model which support AP signal propagation where the model is written in the form of PDE instead of the usual ODE form (which describes signal evolution in time at a fixed spatial point). In these, normally, the potential-gradient-type member responsible for propagating the signal along the axon is not as clearly defined from a physical first-principles viewpoint (usually some kind of abstracted diffusion-type process is used). This is essential later for the purpose of clearer physical interpretation as we make use of the model based on the elementary form of Maxwell equations which is further modified to include the influence of myelination on the signal propagating along the axon. 

The model \eqref{LIB1}, \eqref{LIB2} based on Maxwell equations for a transmission line can be modified to include the influence of myelination. In addition to the g-ratio normally considered in the earlier studies (taken into account indirectly through coefficient $\gamma$) which takes into account the myelination geometry perpendicular to the axon we introduce the so-called ``myelination-ratio" or $\mu$-ratio which describes the influence of myelin distribution on the signal propagation in the direction of the axis of the axon. The numerical solutions, using parameters from the literature, demonstrate physiologically plausible behaviour for the model. The model is reduced to the model of the unmyelinated axon if the length of the myelinated sections along the axon is taken as zero. Under the considered parameter combinations we can observe the AP propagation velocities up to 67.7 $[\mathrm{m/s}]$ \cite{Tamm2024} (the signal propagation velocity range for myelinated axons is given as roughly 10 to 120 $[\mathrm{m/s}]$ in the earlier studies \cite{Lodish2004,Schmidt2019}).

It is important to emphasise that the proposed continuum-based model is philosophically similar to how the transmission line equations are composed. The `unit-cell' in the context of the myelinated axon in the model is composed of the node of Ranvier and the myelinated section next to it. This is opposed to the alternative approach which is also relatively popular in the literature where the classical HH model is used in the nodes of Ranvier while myelinated sections are handled separately either through some numerical scheme or by an alternative model coupled with the HH model in the node of Ranvier through some mechanism. Having a relatively simple pair of PDEs which are connected to the fundamental principles in physics (i.e., Maxwell equations for anything involving the movement of charges in an environment) could be considered superior to investigating causal connections and making predictions than something that is not as clearly connected to the first principles of physics. 

AP in a myelinated axon based on \cite{Tamm2024} is birefly summarized here. In unmyelinated case we can construct governing equations by drawing inspiration from previous work by Lieberstein \cite{Lieberstein1967} as
\begin{equation} \label{LIB1}
\left(C_a \pi a^2 + C_m 2 \pi a\right)\frac{\partial Z}{\partial t} + \frac{\partial i_a}{\partial x} + 2 \pi a \cdot \left[\hat{g_K} n^4 (Z-Z_K) + \hat{g_{Na}} m^3 h (Z-Z_{Na}) + \hat{g_l} (Z-Z_l)  \right] = 0,
\end{equation}
and 
\begin{equation} \label{LIB2}
\frac{L}{\pi a^2}\frac{\partial i_a}{\partial t} + \frac{\partial Z}{\partial x} + r i_a = 0.
\end{equation}
Here  
 $x$ is space (length) and $t$ is time,
 $Z$ is the action potential, $i_a$ is the line axon current (along the axon) and $i$ is the membrane current per unit length (taken the same as HH current across the membrane),
 $a$ is the radius of the axon, $r$ is the axon resistance per unit length, $L$ is the axon specific self-inductance, $C_a$ is the axon self capacitance per unit area per unit length and $C_m$ is the membrane capacity per unit area. The internal variables $n,m,h$ controlling the opening and closing the of the ion channels and corresponding conductances $\hat{g_K}$, $\hat{g_{Na}}$, $\hat{g_l}$ are taken as outlined in the classical Hodgkin-Huxley model \cite{Hodgkin1952}.
 
Taking the elementary form of Maxwell equations combined with the ideas proposed by Lieberstein in eqs.~\eqref{LIB1} and \eqref{LIB2} as a starting point, we proceed to modify these governing equations to include the effect of myelination on the AP signal propagation.
When we modify the Lieberstein model \cite{Lieberstein1967} to account for the effect of myelination on a nerve fibre, we consider the following hypotheses:
(i) The velocity of the AP depends on the ratio of lengths between the myelin sheath and the node of Ranvier $\left( L_2/L_1 \right)$ (so-called `$\mu$-ratio' below);
(ii) The thickness of the myelin sheath affects the velocity of the AP signal (the so-called $g$-ratio) and could be taken into account indirectly through the capacitance variations (included in parameter $\gamma$ below); 
(iii) The dominant mechanism through which the AP signal velocity in myelinated nerve fibre is increased is the so-called saltatory conduction hypothesis \cite{Bressloff2014};
(iv) The model equation should be reduced back to the basic model when the myelination approaches to zero (i.e., unmyelinated axon). 

Let us take Lieberstein eqs.~\eqref{LIB1} and \eqref{LIB2}, introducing parameters $\mu$ and $\gamma$ characterizing the AP propagation velocity increase from saltatory conduction \cite{Bressloff2014} and other relevant mechanisms. 
Note that Bressloff \cite{Bressloff2014} has used parameter $D$ that modulates the signal dynamics across the membrane. Here, inspired by Bressloff \cite{Bressloff2014}, we have introduced two parameters: $\gamma$ and $\mu$. The governing equations are written in the form:
\begin{equation} \label{LIB1m}
\frac{\partial Z}{\partial t} + \Phi \cdot \left[ \left( 1 + \gamma \cdot \mu\right) \cdot \frac{\partial i_a}{\partial x} + 2 \pi a \cdot V \right] =0,
\end{equation}
\begin{equation} \label{LIB2m}
\frac{\partial i_a}{\partial t} + \frac{\pi a^2}{L} \cdot \left[\frac{\partial Z}{\partial x} + r i_a \right]=0,
\end{equation}
\begin{equation} \label{LIB3m}
\Phi=\frac{1}{C_a \pi a^2 + 2 C_m \pi a},
\end{equation}
\begin{equation} \label{LIB4m}
\mu = \frac{L_2}{L_1}, 
\end{equation}
\begin{equation} \label{LIB5m}
V =  \left(\hat{g_K} n^4 (Z-Z_K) + \hat{g_{Na}} m^3 h (Z-Z_{Na}) + \hat{g_l} (Z-Z_l)  \right) .
\end{equation}
In eq.~\eqref{LIB1m} parameter $\mu$ (describing $\mu$-ratio) describes the average length of the myelinated section ($L_2$) divided by the average length of the node of Ranvier ($L_1$) (see Fig.~\ref{fig4}). It affects the quantity $i_a$ which is the current along the axis of the axon. Parameter $\gamma$ is a phenomenological coefficient which determines conduction velocity between adjacent nodes of Ranvier. Here parameter $\gamma$ includes myelin geometry perpendicular to the axon (related to $g$-ratio, parameters $d_1$ and $d_2$ in Fig.~\ref{fig4}). It should be emphasized that parameter $\gamma$ is not the same as proposed  by Bressloff \cite{Bressloff2014} and is a generalised quantity here.

\end{appendices}


\begin{thebibliography}{10}

\bibitem{Ablowitz2011}
M.~J. Ablowitz.
\newblock {\em {Nonlinear Dispersive Waves}}.
\newblock Cambridge University Press, Cambridge, 2011.

\bibitem{Andersen2009}
S.~S. Andersen, A.~D. Jackson, and T.~Heimburg.
\newblock {Towards a thermodynamic theory of nerve pulse propagation}.
\newblock {\em Prog. Neurobiol.}, 88(2):104--113, 2009.

\bibitem{Bassingthwaighte1995}
J.~B. Bassingthwaighte.
\newblock {Toward Modeling the Human Physionome}.
\newblock pages 331--339. 1995.

\bibitem{Bialek2018}
W.~Bialek.
\newblock {Perspectives on theory at the interface of physics and biology}.
\newblock {\em Reports Prog. Phys.}, 81(1):012601, 2018.

\bibitem{Bressloff2014}
P.~C. Bressloff.
\newblock {\em {Waves in Neural Media}}.
\newblock Lecture Notes on Mathematical Modelling in the Life Sciences.
  Springer New York, New York, NY, 2014.

\bibitem{Carrillo2023}
N.~Carrillo and S.~Mart{\'{i}}nez.
\newblock {Scientific Inquiry: From Metaphors to Abstraction}.
\newblock {\em Perspect. Sci.}, 31(2):233--261, 2023.

\bibitem{Chen2019}
H.~Chen, D.~Garcia-Gonzalez, and A.~J{\'{e}}rusalem.
\newblock {Computational model of the mechanoelectrophysiological coupling in
  axons with application to neuromodulation}.
\newblock {\em Phys. Rev. E}, 99(3):032406, 2019.

\bibitem{Clay2005}
J.~R. Clay.
\newblock {Axonal excitability revisited}.
\newblock {\em Prog. Biophys. Mol. Biol.}, 88(1):59--90, 2005.

\bibitem{Debanne2011}
D.~Debanne, E.~Campanac, A.~Bialowas, E.~Carlier, and G.~Alcaraz.
\newblock {Axon physiology}.
\newblock {\em Physiol. Rev.}, 91(2):555--602, 2011.

\bibitem{DeLanda2002}
M.~DeLanda.
\newblock {\em {Intensive Science and Virtual Philosophy}}.
\newblock Continuum, London, 2002.

\bibitem{Deng2017}
B.~Deng.
\newblock {Alternative Models to Hodgkin–Huxley Equations}.
\newblock {\em Bull. Math. Biol.}, 79(6):1390--1411, 2017.

\bibitem{Drukarch2018}
B.~Drukarch, H.~A. Holland, M.~Velichkov, J.~J. Geurts, P.~Voorn, G.~Glas, and
  H.~W. de~Regt.
\newblock {Thinking about the nerve impulse: A critical analysis of the
  electricity-centered conception of nerve excitability}.
\newblock {\em Prog. Neurobiol.}, 169:172--185, 2018.

\bibitem{Drukarch2021}
B.~Drukarch, M.~M.~M. Wilhelmus, and S.~Shrivastava.
\newblock {The thermodynamic theory of action potential propagation: a sound
  basis for unification of the physics of nerve impulses}.
\newblock {\em Rev. Neurosci.}, 33(3):285--302, 2022.

\bibitem{ElHady2015}
A.~{El Hady} and B.~B. Machta.
\newblock {Mechanical surface waves accompany action potential propagation}.
\newblock {\em Nat. Commun.}, 6:6697, 2015.

\bibitem{Engelbrecht2018c}
J.~Engelbrecht, T.~Peets, and K.~Tamm.
\newblock {Electromechanical coupling of waves in nerve fibres}.
\newblock {\em Biomech. Model. Mechanobiol.}, 17(6):1771--1783, 2018.

\bibitem{Engelbrecht2018}
J.~Engelbrecht, T.~Peets, K.~Tamm, M.~Laasmaa, and M.~Vendelin.
\newblock {On the complexity of signal propagation in nerve fibres}.
\newblock {\em Proc. Estonian Acad. Sci.}, 67(1):28--38, 2018.

\bibitem{EngelbrechtTammPeets2014}
J.~Engelbrecht, K.~Tamm, and T.~Peets.
\newblock {On mathematical modelling of solitary pulses in cylindrical
  biomembranes}.
\newblock {\em Biomech. Model. Mechanobiol.}, 14(1):159--167, 2015.

\bibitem{EngelbrechtMEDHYP}
J.~Engelbrecht, K.~Tamm, and T.~Peets.
\newblock {Modeling of complex signals in nerve fibers}.
\newblock {\em Med. Hypotheses}, 120:90--95, 2018.

\bibitem{Engelbrecht2020a}
J.~Engelbrecht, K.~Tamm, and T.~Peets.
\newblock {Internal variables used for describing the signal propagation in
  axons}.
\newblock {\em Contin. Mech. Thermodyn.}, 32(6):1619--1627, 2020.

\bibitem{Engelbrecht2020bmmb}
J.~Engelbrecht, K.~Tamm, and T.~Peets.
\newblock {Modelling of processes in nerve fibres at the interface of
  physiology and mathematics}.
\newblock {\em Biomech. Model. Mechanobiol.}, 19(6):2491--2498, dec 2020.

\bibitem{Engelbrecht2020m}
J.~Engelbrecht, K.~Tamm, and T.~Peets.
\newblock {On mechanisms of electromechanophysiological interactions between
  the components of nerve signals in axons}.
\newblock {\em Proc. Estonian Acad. Sci.}, 69(2):81--96, 2020.

\bibitem{Raamat2021}
J.~Engelbrecht, K.~Tamm, and T.~Peets.
\newblock {\em {Modelling of Complex Signals in Nerves}}.
\newblock Springer International Publishing, Cham, 2021.

\bibitem{Peets2022}
J.~Engelbrecht, K.~Tamm, and T.~Peets.
\newblock {Physics shapes signals in nerves}.
\newblock {\em Eur. Phys. J. Plus}, 137(6):696, 2022.

\bibitem{PhilArtikkel}
J.~Engelbrecht, K.~Tamm, and T.~Peets.
\newblock {Signals in nerves from the philosophical viewpoint}.
\newblock {\em Proc. Estonian Acad. Sci.}, 71(4):369, 2022.

\bibitem{NovaPeatykk}
J.~Engelbrecht, K.~Tamm, and T.~Peets.
\newblock {Axons' Signals}.
\newblock In A.~Costa and E.~Villalba, editors, {\em Horizons in Neuroscience Research. Volume 49}, chapter~3, page 223. Nova Science Publishers, Inc., New York, NY,
  2023.

\bibitem{Engelbrecht2024}
J.~Engelbrecht, K.~Tamm, and T.~Peets.
\newblock {On the phenomenological modelling of physical phenomena}.
\newblock {\em Proc. Estonian Acad. Sci.}, 73(3):264, 2024.

\bibitem{FitzHugh1961}
R.~FitzHugh.
\newblock {Impulses and physiological states in theoretical models of nerve
  membrane}.
\newblock {\em Biophys. J.}, 1(6):445--466, 1961.

\bibitem{Goldman1968}
L.~Goldman and J.~S. Albus.
\newblock {Computation of Impulse Conduction in Myelinated Fibers; Theoretical
  Basis of the Velocity-Diameter Relation}.
\newblock {\em Biophys. J.}, 8(5):596--607, 1968.

\bibitem{Hall1999}
C.~W. Hall.
\newblock {\em {Laws and Models: Science, Engineering, and Technology}}.
\newblock CRC Press, Boca Raton, 1999.

\bibitem{Heelan1991}
P.~A. Heelan.
\newblock {Hermeneutical Phenomenology and the Philosophy of Science}.
\newblock In H.~J. Silverman, editor, {\em Gadamer and Hermeneutics. Science, Culture, Literature.},
  pages 213--228. Routledge, New York, 1991.

\bibitem{HJ2007}
T.~Heimburg and A.~Jackson.
\newblock {On the action potential as a propagating density pulse and the role
  of anesthetics}.
\newblock {\em Biophys. Rev. Lett.}, 02(01):57--78, 2007.

\bibitem{Heimburg2005}
T.~Heimburg and A.~D. Jackson.
\newblock {On soliton propagation in biomembranes and nerves.}
\newblock {\em Proc. Natl. Acad. Sci. USA}, 102(28):9790--9795, 2005.

\bibitem{Hodgkin1964a}
A.~L. Hodgkin.
\newblock {\em {The Conduction of the Nervous Impulse}}.
\newblock Liverpool University Press, 1964.

\bibitem{Hodgkin1964}
A.~L. Hodgkin.
\newblock {The ionic basis of nervous conduction}.
\newblock {\em Science}, 145(3637):1148--1154, 1964.

\bibitem{Hodgkin1952}
A.~L. Hodgkin and A.~F. Huxley.
\newblock {A quantitative description of membrane current and its application
  to conduction and excitation in nerve}.
\newblock {\em J. Physiol.}, 117(4):500--544, 1952.

\bibitem{Holdsworth2006}
D.~Holdsworth.
\newblock {Becoming Interdisciplinary: Making Sense of DeLanda's Reading of
  Deleuze}.
\newblock {\em Paragraph}, 29(2):139--156, 2006.

\bibitem{Holland2019}
L.~Holland, H.~W. de~Regt, and B.~Drukarch.
\newblock {Thinking About the Nerve Impulse: The Prospects for the Development
  of a Comprehensive Account of Nerve Impulse Propagation}.
\newblock {\em Front. Cell. Neurosci.}, 13(208):1--12, 2019.

\bibitem{Huxley1949}
A.~F. Huxley and R.~St{\"{a}}mpfli.
\newblock {Evidence for saltatory conduction in peripheral myelinated nerve
  fibres.}
\newblock {\em J. Physiol.}, 108(3):315--39, 1949.

\bibitem{Iwasa1980}
K.~Iwasa, I.~Tasaki, and R.~Gibbons.
\newblock {Swelling of nerve fibers associated with action potentials}.
\newblock {\em Science}, 210(4467):338--339, 1980.

\bibitem{Jerusalem2019}
A.~Jerusalem, Z.~Al-Rekabi, H.~Chen, A.~Ercole, M.~Malboubi,
  M.~Tamayo-Elizalde, L.~Verhagen, and S.~Contera.
\newblock {Electrophysiological-mechanical coupling in the neuronal membrane
  and its role in ultrasound neuromodulation and general anaesthesia}.
\newblock {\em Acta Biomater.}, 97:116--140, 2019.

\bibitem{Jerusalem2014}
A.~J{\'{e}}rusalem, J.~A. Garc{\'{i}}a-Grajales, A.~Merch{\'{a}}n-P{\'{e}}rez,
  and J.~M. Pe{\~{n}}a.
\newblock {A computational model coupling mechanics and electrophysiology in
  spinal cord injury}.
\newblock {\em Biomech. Model. Mechanobiol.}, 13(4):883--896, 2014.

\bibitem{Kaufmann1989}
K.~Kaufmann.
\newblock {\em {Action Potentials and Electromechanical Coupling in the
  Macroscopic Chiral Phospholipid Bilayer}}.
\newblock Caruaru, 1989.

\bibitem{Lefebvre2024}
A.~T. Lefebvre, C.~L. Rodriguez, E.~Bar-Kochba, N.~E. Steiner, M.~Mirski, and
  D.~W. Blodgett.
\newblock {High-resolution transcranial optical imaging of in vivo neural
  activity}.
\newblock {\em Sci. Rep.}, 14(1):24756, 2024.

\bibitem{Lieberstein1967}
H.~Lieberstein.
\newblock {On the Hodgkin-Huxley partial differential equation}.
\newblock {\em Math. Biosci.}, 1(1):45--69, 1967.

\bibitem{Lillie1925}
R.~S. Lillie.
\newblock {Factors affecting transmission and recovery in the passive iron
  nerve model}.
\newblock {\em J. Gen. Physiol.}, 7(4):473--507, 1925.

\bibitem{Ling2020}
T.~Ling, K.~C. Boyle, V.~Zuckerman, T.~Flores, C.~Ramakrishnan, K.~Deisseroth,
  and D.~Palanker.
\newblock {High-speed interferometric imaging reveals dynamics of neuronal
  deformation during the action potential}.
\newblock {\em Proc. Natl. Acad. Sci.}, page 201920039, 2020.

\bibitem{Lodish2004}
H.~Lodish, A.~Berk, P.~Matsudaira, C.~A. Kaiser, M.~Krieger, and M.~P. Scott.
\newblock {Transport of Ions and Small Molecules Across Cell Membranes}.
\newblock {\em Mol. Cell Biol.}, pages 245--300, 2004.

\bibitem{Malischewsky1987}
P.~Malischewsky.
\newblock {\em {Surface waves and discontinuities}}.
\newblock Elsevier, Amsterdam, 1987.

\bibitem{Maugin1990}
G.~A. Maugin.
\newblock {Internal variables and dissipative structures}.
\newblock {\em J. Non-Equilibrium Thermodyn.}, 15(2), 1990.

\bibitem{McCulloch2002}
A.~D. McCulloch and G.~Huber.
\newblock {Integrative Biological Modelling In Silico}.
\newblock In G.~Bock and J.~A. Goode, editors, {\em ‘In Silico’ Simulation of Biological Processes}, pages 4--25. John Wiley {\&} Sons, Chichester, 2002.

\bibitem{MelvinLieberstein1970}
H.~{Melvin Lieberstein} and M.~A. Mahrous.
\newblock {A source of large inductance and concentrated moving magnetic fields
  on axons}.
\newblock {\em Math. Biosci.}, 7(1-2):41--60, 1970.

\bibitem{Morris1981}
C.~Morris and H.~Lecar.
\newblock {Voltage oscillations in the barnacle giant muscle fiber}.
\newblock {\em Biophys. J.}, 35(1):193--213, 1981.

\bibitem{Morrison1999}
M.~Morrison.
\newblock {Models as autonomous agents}.
\newblock In M.~Morgan and M.~Morrison, editors, {\em Models as Mediators}, pages
  38--65. Cambridge University Press, Cambridge, 1999.

\bibitem{Mueller2014}
J.~K. Mueller and W.~J. Tyler.
\newblock {A quantitative overview of biophysical forces impinging on neural
  function}.
\newblock {\em Phys. Biol.}, 11(5):051001, 2014.

\bibitem{Mussel2021}
M.~Mussel and M.~F. Schneider.
\newblock {Sound pulses in lipid membranes and their potential function in
  biology}.
\newblock {\em Prog. Biophys. Mol. Biol.}, 162:101--110, 2021.

\bibitem{Nagumo1962}
J.~Nagumo, S.~Arimoto, and S.~Yoshizawa.
\newblock {An active pulse transmission line simulating nerve axon}.
\newblock {\em Proc. IRE}, 50(10):2061--2070, 1962.

\bibitem{NationalResearchCouncil2005}
{National Research Council}.
\newblock {\em {Catalyzing Inquiry at the Interface of Computing and Biology}}.
\newblock The National Academies Press, Washington, 2005.

\bibitem{Noble2002}
D.~Noble.
\newblock {Chair's Introduction}.
\newblock In G.~Bock and J.~A. Goode, editors, {\em ‘In Silico’ Simulation of Biological Processes: Novartis Foundation Symposium 247, Volume 247}, pages 1--3. Novartis
  Foundation, 2002.

\bibitem{Noble2002a}
D.~Noble.
\newblock {The rise of computational biology}.
\newblock {\em Nat. Rev. Mol. Cell Biol.}, 3(6):459--463, 2002.

\bibitem{Peets2021}
T.~Peets, K.~Tamm, and J.~Engelbrecht.
\newblock {On the Physical Background of Nerve Pulse Propagation: Heat and
  Energy}.
\newblock {\em J. Non-Equilibrium Thermodyn.}, 46(4):343--353, 2021.

\bibitem{Peets2023}
T.~Peets, K.~Tamm, and J.~Engelbrecht.
\newblock {On mathematical modeling of the propagation of a wave ensemble
  within an individual axon}.
\newblock {\em Front. Cell. Neurosci.}, 17, 2023.

\bibitem{Pennycuick1992}
C.~J. Pennycuick.
\newblock {\em {Newton Rules Biology : A Physical Approach to Biological
  Problems}}.
\newblock Oxford University Press, Oxford, 1992.

\bibitem{Portides2011}
D.~Portides.
\newblock {Seeking representations of phenomena: Phenomenological models}.
\newblock {\em Stud. Hist. Philos. Sci.}, 42(2):334--341, 2011.

\bibitem{Porubov2003}
A.~V. Porubov.
\newblock {\em {Amplification of Nonlinear Strain Waves in Solids}}.
\newblock World Scientific, Singapore, 2003.

\bibitem{Purves2017}
D.~Purves, G.~J. Augustine, D.~Fitzpatrick, W.~C. Hall, A.-S. LaMantia, R.~D.
  Mooney, M.~L. Platt, and L.~E. White.
\newblock {\em {Neuroscience}}.
\newblock Sinauer Associates, New York, 6th editio edition, 2017.

\bibitem{Robinson2018}
A.~Robinson.
\newblock {Did Einstein really say that?}
\newblock {\em Nature}, 557(7703):30--30, 2018.

\bibitem{Rovelli2022}
C.~Rovelli.
\newblock {\em {Helgoland}}.
\newblock Penguin Random House, London, 2022.

\bibitem{Rvachev2010}
M.~M. Rvachev.
\newblock {On axoplasmic pressure waves and their possible role in nerve
  impulse propagation}.
\newblock {\em Biophys. Rev. Lett.}, 5(2):73--88, 2010.

\bibitem{Schmidt2019}
H.~Schmidt and T.~R. Kn{\"{o}}sche.
\newblock {Action potential propagation and synchronisation in myelinated
  axons}.
\newblock {\em PLOS Comput. Biol.}, 15(10):e1007004, 2019.

\bibitem{Schneider2021}
M.~F. Schneider.
\newblock {Living systems approached from physical principles}.
\newblock {\em Prog. Biophys. Mol. Biol.}, 162:2--25, 2021.

\bibitem{Sherrington1907}
C.~S. Sherrington.
\newblock {The Integrative Action of the Nervous System}.
\newblock {\em Nature}, 76(1962), 1907.

\bibitem{Tamm2019}
K.~Tamm, J.~Engelbrecht, and T.~Peets.
\newblock {Temperature changes accompanying signal propagation in axons}.
\newblock {\em J. Non-Equilibrium Thermodyn.}, 44(3):277--284, 2019.

\bibitem{Tamm2015}
K.~Tamm and T.~Peets.
\newblock {On solitary waves in case of amplitude-dependent nonlinearity}.
\newblock {\em Chaos, Solitons {\&} Fractals}, 73:108--114, 2015.

\bibitem{Tamm2021}
K.~Tamm, T.~Peets, and J.~Engelbrecht.
\newblock {Mechanical waves in myelinated axons}.
\newblock {\em Biomech. Model. Mechanobiol.}, 21(4):1285--1297, 2022.

\bibitem{Tamm2024}
K.~Tamm, T.~Peets, and J.~Engelbrecht.
\newblock {The modelling of the action potentials in myelinated nerve fibres}.
\newblock {	arXiv:2406.18590 [physics.bio-ph]}, 2024.

\bibitem{Tasaki1988}
I.~Tasaki.
\newblock {A macromolecular approach to excitation phenomena: mechanical and
  thermal changes in nerve during excitation}.
\newblock {\em Physiol. Chem. Phys. Med. NMR}, 20(4):251--268, 1988.

\bibitem{Tasaki1989}
I.~Tasaki, K.~Kusano, and P.~M. Byrne.
\newblock {Rapid mechanical and thermal changes in the garfish olfactory nerve
  associated with a propagated impulse}.
\newblock {\em Biophys. J.}, 55(6):1033--1040, 1989.

\bibitem{Terakawa1985}
S.~Terakawa.
\newblock {Potential-dependent variations of the intracellular pressure in the
  intracellularly perfused squid giant axon.}
\newblock {\em J. Physiol.}, 369(1):229--248, 1985.

\bibitem{Tomassy2014}
G.~S. Tomassy, D.~R. Berger, H.-H. Chen, N.~Kasthuri, K.~J. Hayworth,
  A.~Vercelli, H.~S. Seung, J.~W. Lichtman, and P.~Arlotta.
\newblock {Distinct Profiles of Myelin Distribution Along Single Axons of
  Pyramidal Neurons in the Neocortex}.
\newblock {\em Science}, 344(6181):319--324, 2014.

\bibitem{Truesdell1965}
C.~Truesdell and W.~Noll.
\newblock {\em {The non-linear field theories of mechanics}}.
\newblock Springer, Berlin, 1965.

\bibitem{Wang2021}
H.~Wang, J.~Wang, G.~Cai, Y.~Liu, Y.~Qu, and T.~Wu.
\newblock {A Physical Perspective to the Inductive Function of Myelin—A
  Missing Piece of Neuroscience}.
\newblock {\em Front. Neural Circuits}, 14:1--23, jan 2021.

\bibitem{Whitham1974}
G.~Whitham.
\newblock {\em {Linear and Nonlinear Waves}}.
\newblock Wiley-Interscience, New York, 1974.

\bibitem{Winlow2024}
W.~Winlow.
\newblock {Editorial: 90th anniversary of the 1932 Sherrington and Adrian Nobel
  prize: new insights into initiation and propagation of action potentials and
  behavioural modulation of reflexes}.
\newblock {\em Front. Cell. Neurosci.}, 18, 2024.

\bibitem{Yang2018}
Y.~Yang, X.-W. Liu, H.~Wang, H.~Yu, Y.~Guan, S.~Wang, and N.~Tao.
\newblock {Imaging Action Potential in Single Mammalian Neurons by Tracking the
  Accompanying Sub-Nanometer Mechanical Motion}.
\newblock {\em ACS Nano}, 12(5):4186--4193, 2018.

\end{thebibliography}

\end{document}